\documentclass[twocolumn,twoside,slac_two]{revtex4}
\usepackage{graphicx}
\usepackage{fancyhdr}
\newcommand{\prt}[1]{\ensuremath{\mathrm{#1}}}
\newcommand{\un}[2]{\ensuremath{#1\,\mathrm{#2}}}

\newcommand{\vII}[2]{\ensuremath{%
  \left( \begin{array}{c} #1 \\ #2 \end{array} \right)}}

\newcommand{\mII}[4]{\ensuremath{%
  \left( \begin{array}{cc} #1 & #2 \\ #3 & #4 \end{array} \right)}}

\newcommand{\figref}[1]{Fig~\ref{#1}}

\newcommand{\equiref}[1]{Eq~\ref{#1}}

\newcommand{\secref}[1]{section~\ref{#1}}

\usepackage{color}

\newcommand{\hide}[1]{[]}
\begin{document}

\title{Dalitz Analyses in Charm}

%

\author{Jonas Rademacker on behalf of the CLEO-c Collaboration}
\affiliation{University of Bristol, UK}

\begin{abstract}
  Dalitz analyses in charm give access to magnitudes \emph{and} phases
  of the charm decay amplitudes. They play a significant role in
  charm-mixing measurements, in the measurement of the CP-violating
  phase $\gamma$ in B decays to charm, and in the analysis of light
  meson resonances. We review recent results in all three categories.
\end{abstract}

\maketitle

\thispagestyle{fancy}


\section{Dalitz Analyses}
Dalitz analyses give access to the full complex decay amplitudes,
allowing the measurement of magnitudes \emph{and} phases. This can be
exploited in several ways:
\begin{itemize}
\item Explore the properties of light meson resonances such as the
 \prt{\rho}, \prt{a_1}, \prt{f(980)} and more controversial ones such
 as \prt{\sigma} and \prt{\kappa}.
\item Investigate charm itself, in particular mixing and CP
  violation in charm.
\item B mesons decay to charm most of the time. The phases of the B
  decay amplitude can in certain circumstances be measured by
  analysing the charm Dalitz plot. This provides the theoretically
  cleanest and statistically most powerful direct constraint on the CKM
  phase $\gamma$ that is currently known.
\end{itemize}
In the following we will consider each of these items in turn.
\section{Dalitz Plots}
The kinematics of a 3 body decay \prt{D\to A, B, C} (such as \prt{D^+
  \to K^+ K^- \pi^+}) can be fully described by 2 parameters. In terms
of the four momenta of the three decay products, which we will denote
as $p_A, p_B, p_C$, one usually picks the following
invariant-mass-squared parameters:
\begin{eqnarray}
m^2_{AB} &\equiv& \left(p_A + p_B\right)^2 \\
m^2_{BC} &\equiv& \left(p_B + p_C\right)^2
\end{eqnarray}
These parameters are Lorentz invariant, and the phase space density in
terms of these parameters, $\frac{d^2\Phi}{d \left(m_{AB}^2\right)\,d
  \left(m_{BC}^2\right)}$ is flat, i.e. constant inside the
kinematically allowed limits, and zero outside. A Dalitz
plot~\cite{Dalitz:1953cp} is the decay rate in terms of these or
equivalent variables, displayed in a 2-dimensional plot.  The full
decay rate is given by~\cite{BYCKLING_KAJANTIE}:
\begin{eqnarray}
\lefteqn{\frac{d^2\Gamma}{d \left(m^2_{AB}\right)\, d
    \left(m^2_{BC}\right)}}
\nonumber\\
&=&
\left|a_1 e^{i\delta_1} + a_2 e^{i\delta_2} +\ldots\right|^2
\;
\frac{\pi \sqrt{\lambda}
}{2 m_D^2}
\end{eqnarray}
with $\lambda = \left(m_D^2 - m_A^2 - m_B^2\right)^2 - 4 m_A^2 m_B^2$
within the kinematically allowed limits, and $\lambda = 0$ outside.
In the above expression, $a_i e^{i\delta_i}$ describe complex
contributions to the total decay amplitude. In the simplest case, $a_i
e^{i\delta_i}$ are complex Breit-Wigner distributions (or similar
e.g. the Flatt\'e distribution \cite{Flatte:1976xu}) describing
individual particle resonances, with additional factors taking into
account angular momentum conservation, and form factors
(Blatt-Weisskopf penetration factors \cite{BLATTWEISSKOPF}).  This
so-called isobar model has some shortcomings, the most severe one
being that it violates unitarity, especially in the case of wide,
overlapping resonances.  More complicated models such as the K-matrix
formalism \cite{Wigner:1946zz, AitchisonKMatrix}, which respects
unitarity, may therefore be necessary to adequately describe the
observed data, and to provide a theoretically satisfactory model. The
general consensus - at least amongst experimentalists - appears to be
that the isobar description is adequate for $P$ and $D$ wave
resonances, but not for wide $S$ wave resonances. The adequate
description of $L=0$ decays is one of the main topics of interest in
the next section.

\section{Dalitz Analyses and Light Meson Resonance}
\subsection{The \prt{\kappa}, \prt{\sigma} problem}
The S-wave resonances \prt{\sigma \to \pi^+ \pi^-} and \prt{\kappa \to
  K^+ \pi^-} are needed to describe the data in isobar fits to
\prt{D^+ \to \pi^+ \pi^- \pi^+}, \prt{D^0 \to K_S \pi^- \pi^+} and
\prt{D^+ \to K^- \pi^+ \pi^+}. However, it is unclear if this is
compatible with LASS scattering data, and fits to \prt{D^0 \to K^-
  \pi^+ \pi^0}, \prt{D^0 \to \pi^+ \pi^- \pi^0} do not require the
addition of \prt{\sigma} or \prt{\kappa}. K-matrix models do not
explicitly require \prt{\sigma} or \prt{\kappa}.

A wealth of measurements and interesting information has been
published on this topic. Here, we will only consider two decay channels for
which we have recent results, one
for the \prt{\pi^+\pi^-} $S$ wave (\prt{\sigma}), and one for the
\prt{K^+\pi} $S$ wave (\prt{\kappa}).
\subsection{\prt{D^+ \to \pi^+ \pi^- \pi^+}}
Recent analyses of this channel include E791's analysis using an isobar
fit with a \prt{\sigma} resonance~\cite{Aitala:2000xu} , and FOCUS,
who pioneered the K-matrix approach in this channel, and also analyse
\prt{D_S \to \pi^+ \pi^- \pi^+} in their study~\cite{Link:2003gb}.
The most recent result is by CLEO-c~\cite{Bonvicini:2007tc}, using
$\sim 2600$ signal events. CLEO consider isobar models with different
descriptions of the \prt{f_0(980)} and \prt{\sigma}, and two models
that respect unitarity and chirality, one according to
Schechter~\cite{Schechter:2005we} and another developed by
Achasov~\cite{Achasov:2003xn}.
\begin{figure}
\begin{tabular}{cc}
{
  \begin{minipage}[t]{0.46\columnwidth}
  \includegraphics[width=0.99\textwidth]{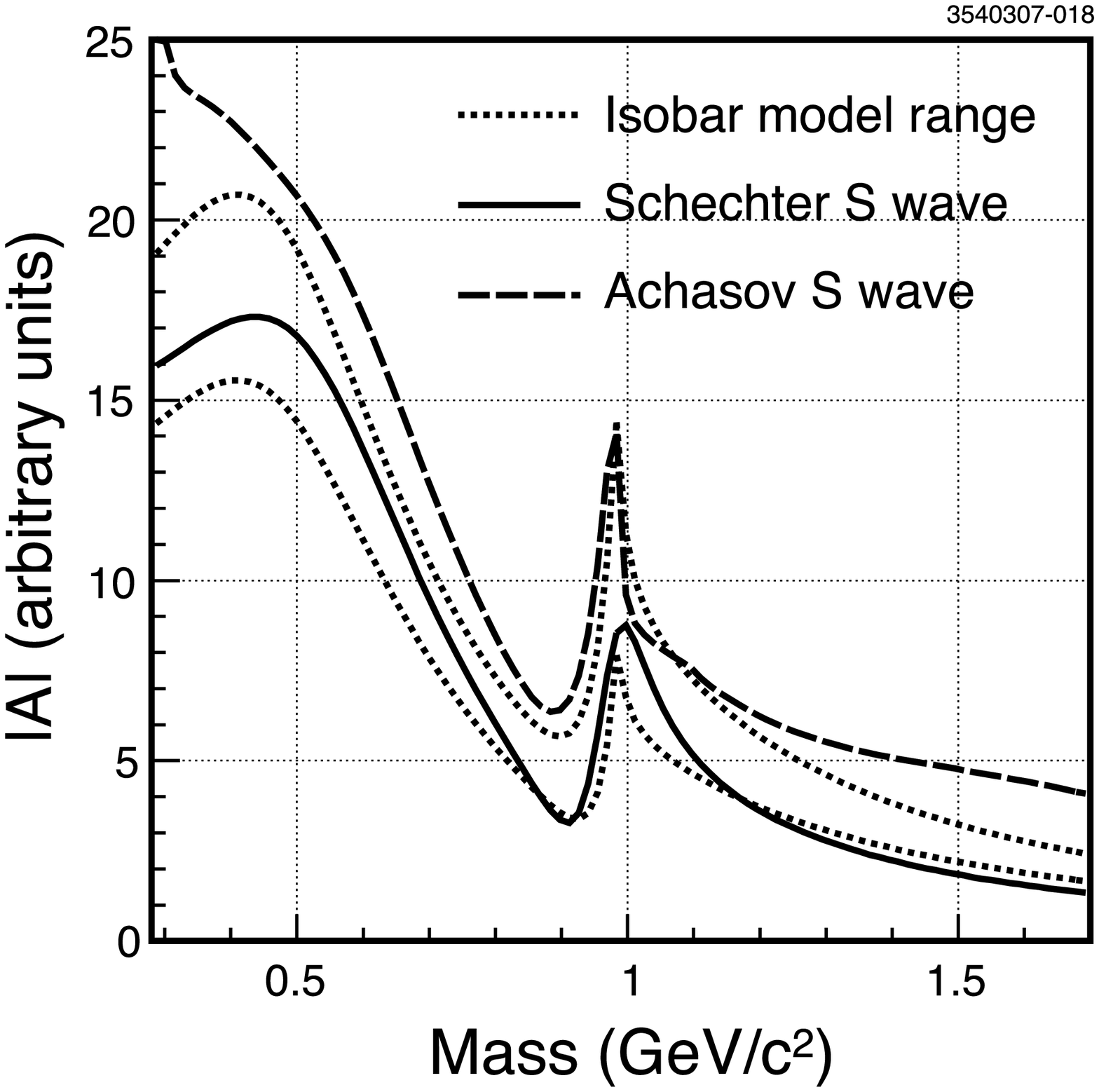}\\
  The $\pi^+\pi^-$ S wave absolute amplitude for different models in
  CLEO-c's \prt{D^0 \to \pi^+\pi^-\pi^+} analysis~\cite{Bonvicini:2007tc}.
  \end{minipage}
} & {
  \begin{minipage}[t]{0.46\columnwidth}
  \includegraphics[width=0.99\textwidth]{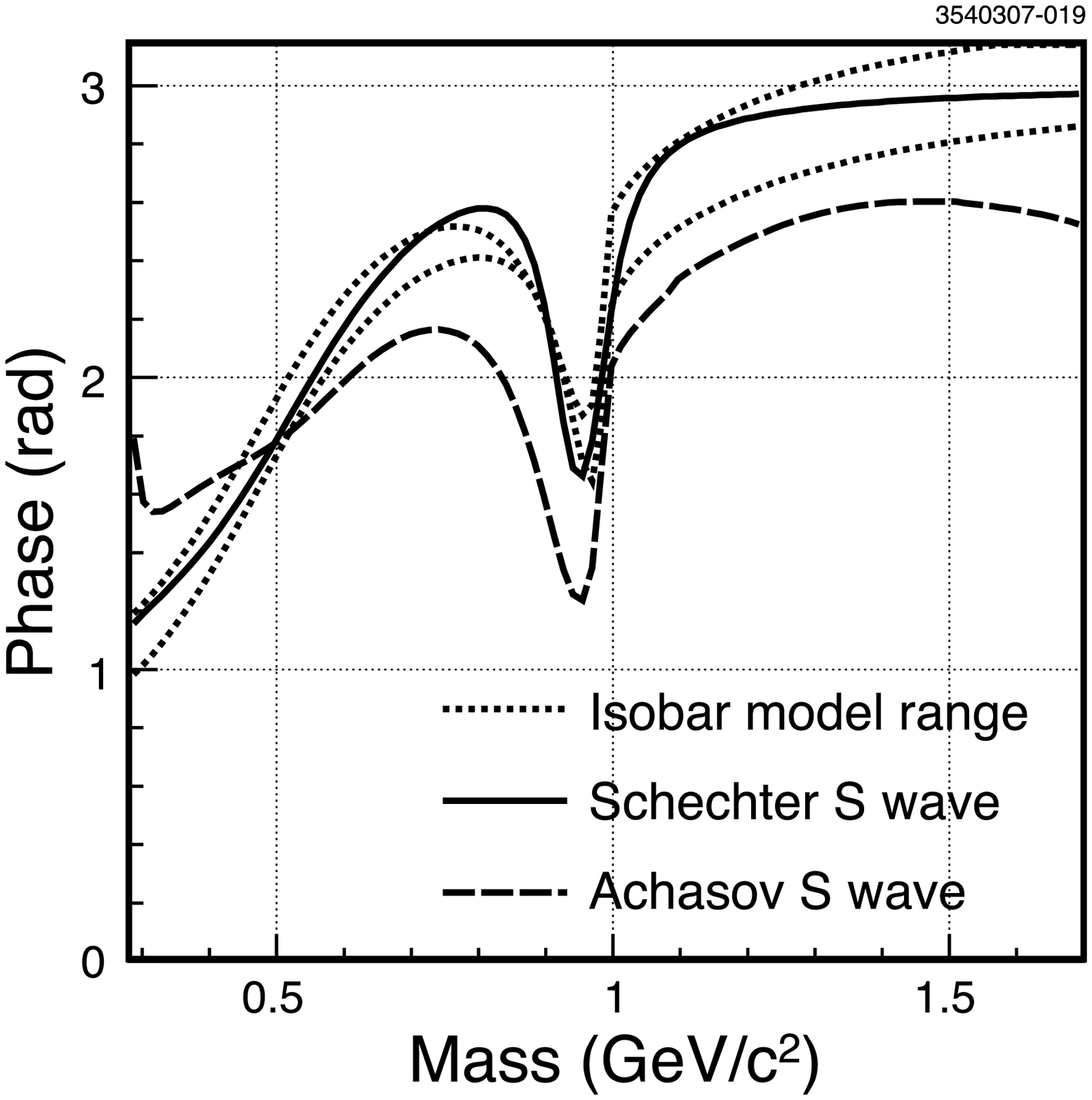}
  \\The $\pi^+\pi^-$ S wave phase for different models in
  CLEO-c's \prt{D^0 \to \pi^+\pi^-\pi^+} analysis~\cite{Bonvicini:2007tc}.
  \end{minipage}
}
\end{tabular}\\
  \caption{ The plots show the magnitude and phase of the amplitude
    of the $S$ wave component for the different models considered in
  CLEO-c's \prt{D^0 \to \pi^+\pi^-\pi^+} analysis~\cite{Bonvicini:2007tc}. The
    dotted lines represent the upper and lower limit of the range of values
    obtained for the various isobar models. The broken line
    corresponds to Achasov's model, the solid line to Schechter's
    model. There is good agreement between the models.
    \label{fig:S_waves_abs_amp}}
\end{figure}
 All models considered agree with each other and the results are
 consistent with previous fits. The amplitude and phase of the $S$
 wave contribution as function of \prt{\pi^+ \pi^-} invariant mass is
 reproduced in figure \figref{fig:S_waves_abs_amp}.

\subsection{\prt{D^+ \to K^- \pi^+ \pi^+}}
\label{sec:DplusKpipipi}
The branching fraction of \prt{D^+ \to K^- \pi^+ \pi^+} is comparably
large, with $\left(9.51 \pm 0.34\right)\%$~\cite{pdg06}. Over 60\% of
its decay rate proceeds via a \prt{K\pi} $S$-wave, as has been
observed in several experiments. In 2002~E791~\cite{Aitala:2002kr},
using an isobar model, found a large \prt{\kappa} contribution. In
2006, E791 re-analysed the same data with a model-independent
description of the $S$ wave, using a binned amplitude and
phase~\cite{Aitala:2005yh}. In 2007, FOCUS~\cite{Pennington:2007se}
applied the K-matrix formalism, constrained by LASS scattering
data~\cite{Aston:1987ir}, to $54k$ events. The most recent result,
which also has the largest data set, is from CLEO-c in 2008, using
140k events, with very little background
($1.1\%$)~\cite{Bonvicini:2008jw}. CLEO-c fit the data using both the
isobar and the model-independent approach, and compare their result to
models used by other experiments. For both types of model, CLEO-c get
a significantly improved fit if they allow for an isospin=2
\prt{\pi^+\pi^-} $S$ wave contribution, where the model-independent
approach gives the better $\chi^2$ per degree of freedom.

\subsection{Four-body ``Dalitz'' Analysis}
Essentially the same formalism as for 3 body decays can be applied to
to 4 body decay. Such analyses are challenging as the equivalent of
the Dalitz plot now has 5 dimensions instead of 2, phase space is not
flat in the usual invariant-mass squared variables, and the amplitude
structure is more complex. A recent example of such an analysis, using
an isobar model, is given by FOCUS for the decay channel \prt{D^0 \to
  \pi^+ \pi^+ \pi^- \pi^-}~\cite{Link:2007fi}. FOCUS observe that
\prt{D^0 \to a_1(1260)\pi} is the dominant decay channel, followed by
\prt{D^0 \to \rho \rho}. The authors find that the \prt{a_1}
predominantly decays to \prt{\sigma \pi}. Many more results can be
found in the paper, including the \prt{\rho \rho} polarisation. Four
body amplitude analyses of \prt{D^0} decays also play as significant
role in extracting $\gamma$ from \prt{B^{\pm} \to D^0 K^{\pm}} decays,
which is discussed in \secref{sec:gamma}.

\section{Charm Mixing and CP violation with Dalitz Plots}
\subsection{The neutral \prt{D^0} system}
The neutral \prt{D} system is the only neutral meson system that mixes
and consists of up-type quarks. It therefore provides a unique window
on Flavour Changing Neutral Currents (FCNC's) between up-type quarks,
which can be affected by new physics in a very different way than
those in down-type quarks, such as investigated in the \prt{B^0_{s,d}}
systems. Particularly interesting is CP violation in the \prt{D}
system, which is essentially zero in the Standard Model, but could be
significant in many New Physics scenarios. The discovery of CP
violation in charm decays would be a clear signal for New Physics.

\paragraph{Mixing Parameters}
The physical mass eigenstates \prt{D_1}, \prt{D_2} of the neutral
\prt{D} meson are superpositions of the flavour-specific states
\prt{D^0} and \prt{\bar{D}^0}:
\begin{eqnarray}
\prt{D_1} & = & p \prt{D^0} + q \prt{\bar{D}^0} \\
\prt{D_2} & = & p \prt{D^0} - q \prt{\bar{D}^0} 
\end{eqnarray}
where $p$ and $q$ are complex numbers satisfying $|p|^2 + |q|^2 =
1$. The mass and the width difference between \prt{D_1} and \prt{D_2}
are $\Delta m$ and $\Delta \Gamma$. Mixing in the neutral \prt{D}
system is conventionally parametrised by the parameters $x$ and $y$
given by
\begin{eqnarray}
x & = & \frac{\Delta m}{\bar{\Gamma}} \\
y & = & \frac{\Delta \Gamma}{\bar{\Gamma}}
\end{eqnarray}
In the absence of mixing, both parameters are zero. The Standard model
expectations for $x$ and $y$ are $\sim 10^{-3} -
10^{-2}$~\cite{DmixSMPrediction}.

\paragraph{CP Violation in Charm}
If $\left|\frac{q}{p}\right| \neq 1$, CP is violated (CP violation in
mixing). CP violation in the interference between mixing and decay is
parametrised by the phase $\phi$. The phase $\phi$ is the exact
equivalent in the neutral \prt{D^0} system of the parameter $-2\beta$
in the neutral \prt{B^0_d} system measured in \prt{B^0 \to J/\psi K_s}
by the B factories. The phase of the ratio $\frac{q}{p}$ is convention
dependent and without fixing the convention, its value does not say
anything about CP violation. Usually, however, a convention is chosen
where $\frac{q}{p} = \left|\frac{q}{p}\right| e^{i\phi}$, so that both
kinds of CP violation discussed here are encoded in the same complex
ratio $q/p$.

In the SM, CP violation in charm is zero for all practical purposes,
i.e. relative to the sensitivities of current experiments and those
planned for the foreseeable future. However, many NP models predict CP
violation in charm at a level that would be experimentally accessible.
A review with further details on charm mixing, CP violation and its
sensitivity to physics beyond the Standard Model can be found
in~\cite{pdg06}.

\subsection{D mixing using Dalitz Plots}
Evidence for charm mixing has been observed first using the
``wrong-sign'' decay \prt{D^0 \to K^+\pi^-} \cite{Zhang:2006dp,
  Aubert:2007wf, cdfDmix:2007uc}. The final decay rate has
contributions from the DCS amplitude \prt{D^0 \to K^+ \pi^-} and a
combination of mixing and a CF amplitude, \prt{D^0 \to \bar{D}^0 \to
  K^+ \pi^-}. A time-dependent measurement is sensitive both to CP
violation and to the mixing parameters ${x^{\prime}}^2$ and
$y^{\prime}$. The primed parameters are related to $x, y$:
\begin{equation}
\vII{x^{\prime}}{y^{\prime}} = 
\mII{\mbox{}\;\;\cos\delta_{K\pi}}{\sin\delta_{K\pi}}
   {-\sin\delta_{K\pi}}{\cos\delta_{K\pi}} \vII{x}{y}
\end{equation}
where $\delta_{K\pi}$ is the phase difference between the amplitude
\prt{D^0 \to K^+\pi^-} and \prt{\bar{D}^0 \to K^+\pi^-}, measured at
CLEO-c to be
\(
 \cos\delta_{K\pi} = 0.9\pm 0.3
\)~\cite{Sun:2007fh}. 
\subsubsection{D mixing and CPV in \prt{D\to K_s \pi \pi}}
CLEO-c pioneered the Dalitz plot analysis of the self-conjugate decay
\prt{D^0 \to K_s \pi^+\pi^-} for the \prt{D^0} mixing
analysis~\cite{Asner:2005sz}. In this case, CF modes such as \prt{D^0
  \to K^{*-}\pi^+} and DCF modes like \prt{D^0 \to K^{*+}\pi^-}, which
contribute to the mixing measurement in a similar way as in the
two-body case, are in the same Dalitz plot, so their relative phase
can be measured. The method gives direct access to $x$ and $y$, and is
also sensitive to the CP violation parameters $|p/q|$ and $\phi$. The
BELLE collaboration analysed approximately \un{0.5}{M}
\prt{D^0,\bar{D^0} \to K_S\pi^+\pi^-} events, with the
result~\cite{Abe:2007rd}:
\begin{eqnarray*}
 x & = & \left(0.81 \pm 0.20^{+0.13}_{-0.17}\right)\% \\
 y & = & \left(0.37 \pm 0.25^{+0.10}_{-0.15}\right)\% \\
 |p/q| & = & 0.86 \pm 0.30^{+0.10}_{-0.09} \\
 \phi  & = & -14^o \pm 18^o \pm 5^o
\end{eqnarray*}
where the first error is statistical and the second error is
systematic.  This measurement dominates the world precision on $x$
\cite{Barberio:2008fa}.

\subsubsection{CPV measurements in SCS decays}
The statistically dominant measurement of CP violation in charm from
\prt{D^0 \to K\pi} and the Dalitz plot method with \prt{D^0 \to K_s
  \pi \pi} study CP violation in amplitudes where doubly Cabibbo
suppressed amplitudes interfere with \prt{D^0} mixing and a subsequent
Cabibbo favoured decay. However, decays involving singly Cabibbo
suppressed decays could in principle be affected by New Physics in a
different way and hence show CP violation even though CF and DCS modes
do not.

\paragraph{\prt{D^0 \to \pi^+ \pi^- \pi^0} and  \prt{D^0 \to K^+ K^- \pi^0}}
BaBar measure several CP violation observable the 3-body decays
\prt{D^0/\bar{D}^0 \to \pi^+ \pi^- \pi^0} and \prt{D^0/\bar{D}^0 \to
  K^+ K^- \pi^0}~\cite{Aubert:2008yd}. BaBar perform a model
independent binned analysis of the rate of \prt{D^0 \to \pi^+ \pi^-
  \pi^0} vs \prt{\bar{D}^0 \to \pi^+ \pi^- \pi^0} across the Dalitz
plot, as well as a model-dependent one, comparing magnitudes and
phases of the CP-conjugate amplitudes. No significant CP asymmetry has
been found. The authors conclude that any CP violation in the singly
Cabibbo-suppressed charm decays occurs at a rate which is not larger
than a few percent.

BELLE consider the asymmetry of the total decay rate in
\prt{D^0/\bar{D}^0 \to \pi^+ \pi^- \pi^0}, integrated across the whole
Dalitz space). Their result~\cite{Arinstein:2008zh},
\(
A_{CP} = (0.43 +/- 1.30)\%,
\)
also shows no evidence for CP violation in SCS decays.

\paragraph{\prt{D^+ \to K^+ K^- \pi^+}}
CLEO-c fit an isobar amplitude model to the Dalitz plot of the SCS
decay \prt{D^+ \to K^+ K^- \pi^+} and its CP conjugate, \prt{D^- \to
  K^- K^+ \pi^-}, and form the asymmetry of the decay fractions
($\propto |A|^2$) for each of the amplitude contributions. No evidence
for CP violation is observed.

\section{Dalitz Analysis in Charm for Precision B physics}
\label{sec:gamma}
\subsection{Introduction}
A central aim of current and future flavour physics experiments is the
precision determination of the CP-violating phase $\gamma$. In terms
of the elements of the Cabibbo-Kobayashi-Maskawa (CKM) quark mixing
matrix, $\gamma$ is defined as
$\mathrm{arg}(-V^\ast_{ub}V_{ud}/V_{cb}^\ast V_{cd})$.

A theoretically clean and statistically powerful method exploits the
interference of \prt{B^{\pm} \to D^0 K^{\pm}} and \prt{B^{\pm} \to
  \bar{D}^0 K^{\pm}} decays, where the \prt{D^0} and \prt{\bar{D}^0}
decay to a common final state $f$~\cite{GLW1, GLW2, ADS, Giri:2003ty,
  Poluektov:2004mf}.  Suitable final states $f$ include 2-body states
such as \prt{KK}, \prt{\pi\pi}~\cite{GLW1, GLW2}, \prt{K\pi}~\cite{ADS},
3-body final states such as \prt{K_S \pi^+ \pi^-} and
\prt{K\pi\pi^0}~\cite{Giri:2003ty, Poluektov:2004mf} and 4-body final
states such as \prt{K\pi\pi\pi}~\cite{ADS, Atwood:2003mj} and
\prt{KK\pi\pi}~\cite{Rademacker:2006zx}.

All such measurements are sensitive to the amplitude ratios
\begin{equation}
\label{eq:rbgamma}
\frac{
A({B}^{\pm}\!\rightarrow {\bar{D}}^{0} {K}^{\pm})}
{A({B}^{\pm}\!\rightarrow {D}^{0} {K}^{\pm})} = r_{B}e^{i(\delta_{B}
  \pm \gamma)}.
\end{equation}
In all cases, the measurement is affected the properties (especially
phases) of the \prt{D^0} decay amplitudes. This is where charm physics
can make a significant contribution to precision B physics. By
measuring the phases of the charm decay amplitudes, the uncertainties
in the $\gamma$ extraction in B physics experiments can be
significantly reduced.

\subsection{CLEO-c and ${\rm\bf B^{\pm} \to D(K_S \pi\pi)K^{\pm}}$}
The best direct constraints on $\gamma$ come from measurements in
\prt{B^{\pm} \to D(K_S \pi\pi)K^{\pm}} and related modes, at the B
factories~\cite{Aubert:2008bd, BELLEgamma:2008wya}. The combined
result is $\gamma = {67^o}^{+32^o}_{-25^o}$~\cite{CKMFitter}. The
dominant systematic uncertainty in these measurements is the model
uncertainties in the description of the \prt{D^0} decay amplitude,
currently between $5^o$ and $9^o$, which would soon limit the
precision of this measurement at the next-generation flavour physics
experiment, LHCb.

CLEO-c's quantum correlated \prt{D\bar{D}} pairs give
model-independent access to both magnitude \emph{and} phase
information of the decay amplitude across the Dalitz plot. This
additional information can be used as input for a model-independent
extraction of $\gamma$ from a binned Dalitz plot
analysis~\cite{Giri:2003ty, Bondar:2005ki, Bondar:2007ir,
  Atwood:2003mj}, thus eliminating the model-uncertainty.

As Dalitz plot variables, we use the invariant-mass squared of the
\prt{K_S \pi^-} and the \prt{K_S \pi^+} pairs, which we denote as
$s_-$ and $s_+$ respectively.  The phase of the \prt{D} decay
amplitude at a given point in Dalitz space is \(
\delta^{K_S\pi\pi}(s_{-}, s_{+})\).  For the phase-difference
between the \prt{D\to K_S \pi \pi} amplitude and the \prt{\bar{D} \to
  K_S \pi \pi} amplitude at the same point in Dalitz space, we define
\begin{equation}
\label{eq:defDelta}
\Delta_{\delta}(s_{-}, s_{+})  \equiv 
\delta^{K_S\pi\pi}(s_{-}, s_{+}) - \delta^{K_S\pi\pi}(s_{+}, s_{-})
\end{equation}
The quantities measured by CLEO-c that provide the input to the
$\gamma$ analyses of the B-factories and LHCb, are the averages of
$\cos\Delta_{\delta}^{K_S\pi\pi}$ and
$\sin\Delta_{\delta}^{K_S\pi\pi}$ for each bin, $c_i$ and $s_i$:
\begin{eqnarray}
c_i & \equiv & \langle \cos\Delta_{\delta} \rangle_i \\
s_i & \equiv & \langle \sin\Delta_{\delta} \rangle_i
\end{eqnarray}
where the index $i$ denotes the $i$th bin.
The analysis of \prt{K_L \pi\pi} events in the similar way to
\prt{K_S\pi\pi}, provides further input to the $c_i$ and $s_i$
measurement in the \prt{K_S\pi\pi} Dalitz plot. The clean environment
at CLEO-c allows the \prt{K_L} reconstruction from kinematic
constraints with high purity.
\begin{figure}
\centering
\includegraphics[width=0.9\columnwidth]{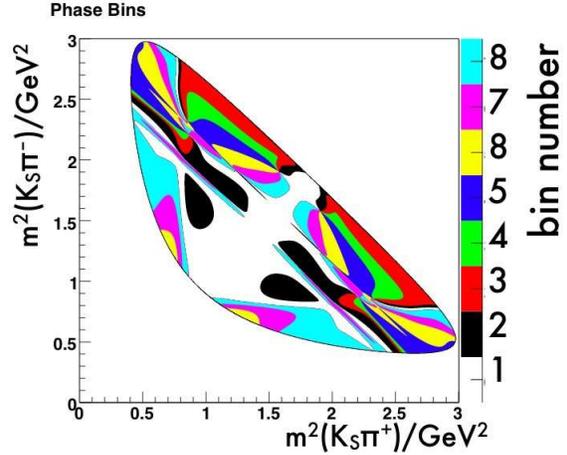}
\caption{Binning used for this preliminary CLEO-c result. The binning
  is uniform in $|\Delta_{\delta}|$, with bin $1$ centred at
  $0^o$.\label{fig:bin}}
\end{figure}
The choice of binning affects the statistical precision of the
analysis - it is beneficial to choose the bins such that the phase
difference $\Delta_{\delta}$, defined in \equiref{eq:defDelta}, varies as
little as possible across each bin~\cite{Bondar:2007ir}. The binning
used for the preliminary CLEO-c results presented here is based on the
BaBar isobar model~\cite{Aubert:2006am}. A uniform binning in
$\Delta_{\delta}$, with eight pairs of bins (arranged symmetrically with
respect to the diagonal axis defined by $s_- = s_+$) is chosen. This
binning is shown in \figref{fig:bin}.
\begin{table}
\begin{tabular}{ || c || *{2}{r@{$\pm$}c@{$\pm$}l|}| }
\hline\hline
 bin number & \multicolumn{3}{c|}{$c_i$} & \multicolumn{3}{c||}{$s_i$}
 \\
\hline\hline
1 &
0.742 & 0.041 & 0.022 & 
-0.022 & 0.168 & 0.096 \\\hline
2 &
0.607& 0.073 & 0.038 &
0.009 & 0.220 & 0.054 \\\hline
3 & 
0.064 & 0.077 & 0.052 &
0.548 & 0.198 & 0.096 \\\hline
4 & 
-0.492 & 0.133 & 0.056 &
0.124 & 0.227 & 0.074 \\\hline
5 &
-0.918 & 0.053 & 0.039 & 
-0.118 & 0.194 & 0.057 \\\hline
6 & 
-0.743 & 0.071 & 0.033 &
-0.296 & 0.203 & 0.070 \\\hline
7 & 
0.156 & 0.092 & 0.050 &
-0.870 & 0.183 & 0.062 \\\hline
8 & 
0.398 & 0.047 & 0.020 &
-0.438 & 0.146 & 0.041 \\\hline\hline
\end{tabular}
\caption{Preliminary CLEO-c results for the measurement of $c_i$ and $s_i$ in
  the \prt{D \to K_S \pi \pi} Dalitz plot, using input from both
  \prt{D \to K_S \pi \pi} events and \prt{D \to K_L \pi \pi}
  events in \un{818}{pb^{-1}}. The first error is the combination of the
  statistical error and the uncertainty that arises from
  the use of \prt{K_L \pi \pi} results for the \prt{K_S \pi \pi} $c_i$
  and $s_i$ determination. The 2nd error is the remaining systematic
  uncertainty. \label{tab:cisi}}
\end{table}
The latest preliminary CLEO results for $c_i$ and $s_i$ from the
combined analysis in both \prt{K_S\pi\pi} and \prt{K_L\pi\pi} Dalitz
plots are shown in Table~\ref{tab:cisi}. When used as input to the
$\gamma$ extraction in the \prt{K_S\pi\pi} mode at the B factories and
LHCb, this is expected to replace the current model uncertainty of
$7^o - 9^o$ with an uncertainty due to the statistically dominated
error on $c_i$ and $s_i$ of about $1^o - 2^o$~\cite{lhcb2007141}.

\subsection{Coherence Factor}
The decay rates \prt{B^{\pm}\to D(hh')K^{\pm}}, where \prt{hh'} stands
for any two-body final state accessible to both \prt{D^0} and
\prt{\bar{D}^0}, are sensitive to $\gamma$~\cite{GLW1, GLW2, ADS}. For
example for \prt{D^0 \to K^+ \pi^-}:
\\\parbox{1.0\columnwidth}{
\begin{eqnarray}
\lefteqn{\Gamma ({ B^{-}\!\to (K^{+}\pi^{-})_{D}K^{-}}) \propto}
\nonumber\\ & & r_B^2  +  {(r_D^{ K\pi})}^2  
{}  {}  +  2  r_B  r_D^{ K\pi}  \cos \left( \delta_B  +  \delta_D^{ K \pi}  -  \gamma \right),\;\;\;\mbox{}
\label{eq:dis1}
\end{eqnarray}
}\\
where we used, in analogy to \equiref{eq:rbgamma}
\begin{equation}
\label{eq:rd}
\frac{
A\left(\prt{D^0 \to K^+ \pi^-}\right)}{
A\left(\prt{\bar{D}^0 \to K^+ \pi^-}\right)}
 = r_{D}^{K\pi}e^{i(\delta_{D}^{K\pi})}.
\end{equation}
The addition of 3 and 4-body decay modes of the D such as \prt{D^0 \to
  K^+ \pi^- \pi^0} and \prt{D^0 \to K^+\pi^-\pi^+\pi^-} can
significantly improve this measurement. In multibody decays \prt{D^o
  \to {\it f}}, the resonant substructure needs to be taken into account,
which can be achieved by adding only one additional parameter to
describe the decay rate~\cite{Atwood:2003mj}. For a generic final
state \prt{{\it f}}:
\begin{eqnarray}
\lefteqn{\Gamma ({ B^{-}} \to (\bar{f})_{ D}{ K^{-}}) \propto}
\nonumber\\ &&
 \bar{A}_{f}^2 + r_B^{2}A_{f}^{2} + 2r_BR_{f}A_{f}\bar{A}_{f}\cos\left( 
\delta_B + \delta_D^{f} - \gamma \right)\;\;\;\;\;\;\mbox{} \label{eq:dis2}
\end{eqnarray}
where $R_{f}$ is the coherence factor which satisfies $0 \le R_{f} \le
1$; $\delta_{D}^{f}$ is the average strong phase difference. The
larger $R_{f}$, the higher the sensitivity to $\gamma$ in a given
mode.  Decays of quantum-correlated \prt{D\bar{D}} pairs at CLEO-c can
be used to measure both $R_f$ and $\delta_D^{f}$.
%
%
\begin{figure}
\begin{tabular}{cc}
\parbox{0.45\columnwidth}{
\mbox{\rotatebox{90}{\mbox{}\hspace{1em}$\delta_D^{K3\pi}$ [deg.]}
\includegraphics[width=0.4\columnwidth]{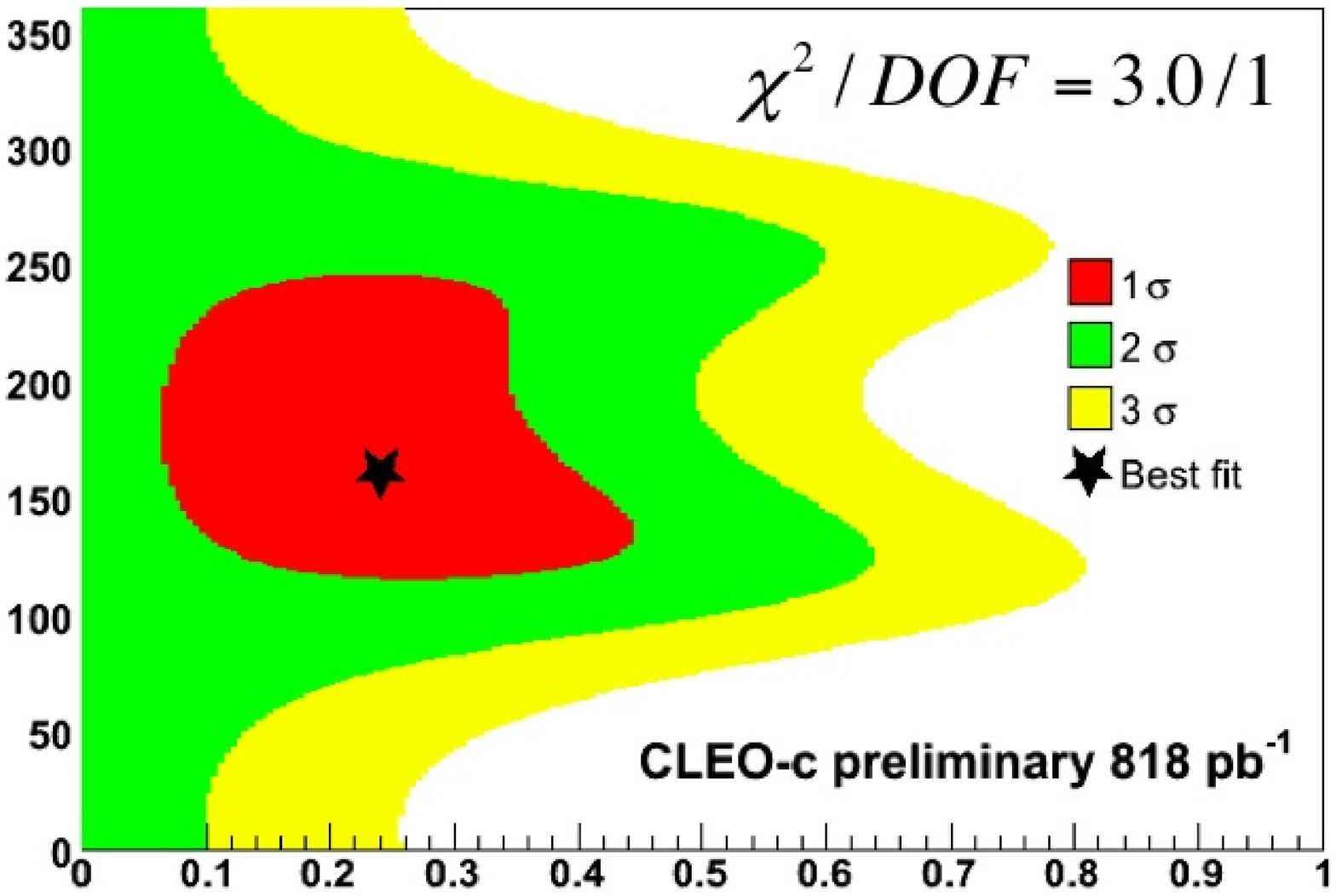}
}\\\mbox{}\hfill $R_{K3\pi}$\hspace{1em}\mbox{}\\}
&
\parbox{0.45\columnwidth}{
\mbox{\rotatebox{90}{\mbox{}\hspace{1em}$\delta_D^{K\pi\pi^0}$ [deg.]}
\includegraphics[width=0.4\columnwidth]{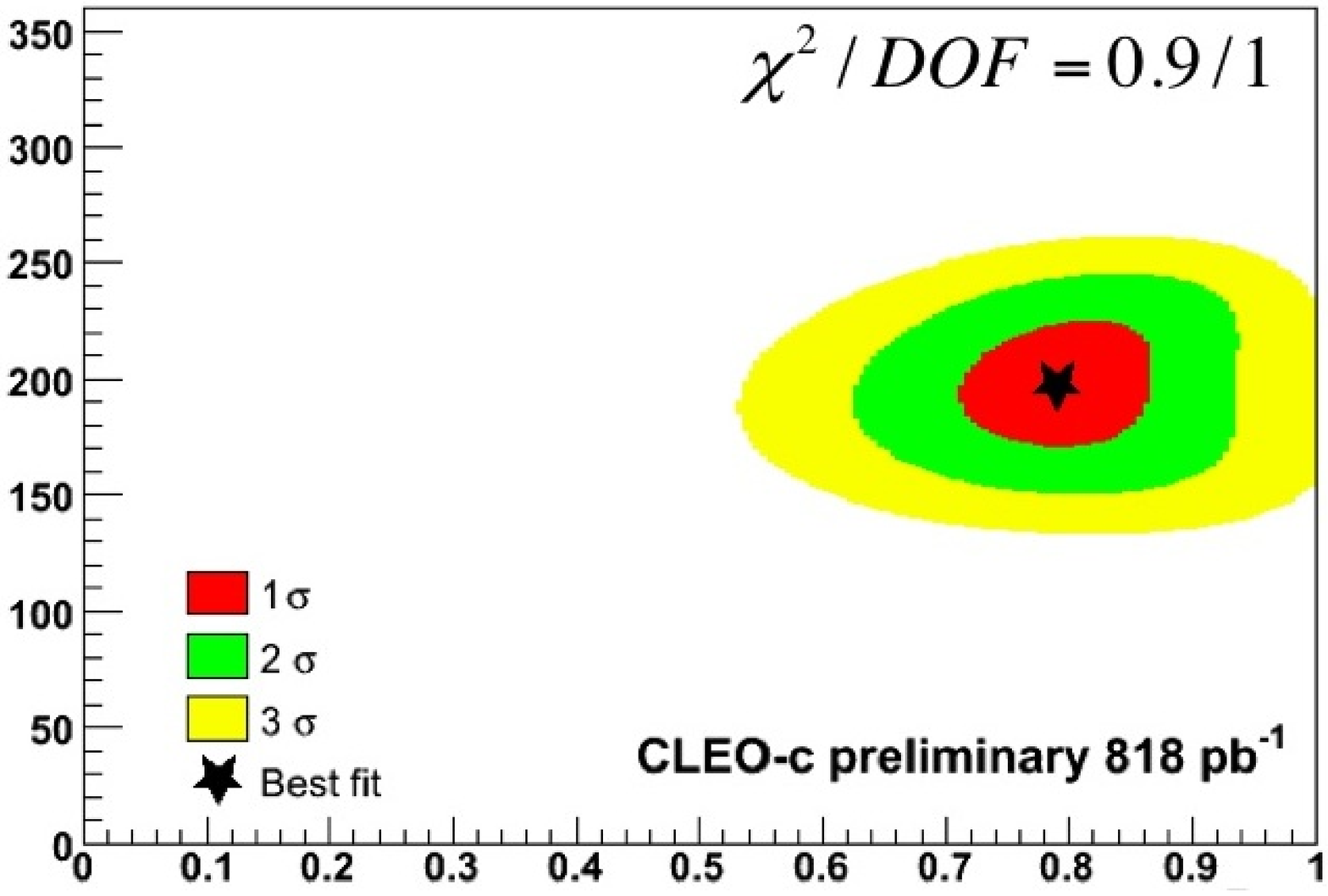}
}\\\mbox{}\hfill $R_{K\pi\pi^0}$\hspace{1em}\mbox{}\\}
\end{tabular}
\caption{1, 2 and 3 $\sigma$ confidence regions in $R_f-\delta_D^f$
  space for \prt{D^0 \to K^+ \pi^- \pi^+ \pi^-} and \prt{D^0 \to K^+
    \pi^- \pi^0}. The star represents the central value. Preliminary
  CLEO-c result, first presented at CKM~2008.\label{fig:ghana}}
\end{figure}
CLEO-c measure the following values for the coherence factors in the
\prt{K^+\pi-\pi^+\pi^-} and \prt{K^+\pi^-\pi^0} mode, as presented for
the first time at CKM~2008~\cite{JimCKM2008}:\\
\begin{tabular}{cc}
\parbox{0.49\columnwidth}{
\begin{eqnarray*}
R_{K3\pi} &=& 0.24^{+0.21}_{-0.17} \\
         &<& 0.57 \;\mbox{at 95\%\ CL}\\
\delta_D^{K3\pi} & = & {161^o}^{+85^o}_{48^o}
\end{eqnarray*}
}
&
\parbox{0.49\columnwidth}{
\begin{eqnarray*}
R_{K\pi\pi^0} &=& 0.79\pm 0.08\\ \\
\delta_D^{K\pi\pi^0} & = & {197^o}^{+28^o}_{27^o}
\end{eqnarray*}
}
\end{tabular}\\
The confidence regions in $R_f$ and $\delta^{f}_D$ space for
\prt{f=K3\pi} and \prt{f=K\pi\pi^0}, as presented at CKM~2008
\cite{JimCKM2008}, are shown in \figref{fig:ghana}.
The small coherence factor in \prt{K3\pi} implies that this mode on
its own would not be very sensitive to $\gamma$. However, this
coherence factor measurement still provides a significant constraint
in a combined measurement with two-body decays of the \prt{D}. The
precise effect depends on the exact value of the strong phase
$\delta^{K\pi}$; typically, without the \prt{K3\pi} constraint from
CLEO-c (but including the $\delta^{K\pi}$ constraint from CLEO-c), the
precision for \un{2}{fb^{-1}} data at LHCb from \prt{D^0 \to KK,
  \pi\pi, K\pi, K3\pi} modes is $\sim 10^o$, which improves to $\sim
8^o$ with the additional information from CLEO-c's coherence factor
measurement in \prt{K3\pi}~\cite{lhcb2007004}. An equivalent study for
\prt{K\pi\pi^0} has not yet been performed, but the large value of
$R_{K\pi\pi^0}$ suggests that we can expect a further significant
improvement.

\section{Conclusion}
Charm physics has recently undergone a remarkable renaissance. Dalitz
analyses in charm give access to magnitudes \emph{and} phases of the
charm decay amplitudes. This can be used to analyse light-meson
resonance, where an improved description of broad S-wave resonances is
of particular interest. Dalitz plots analyses in charm also play a
significant role in charm mixing measurements, where Dalitz analyses
provide the best constraints on $x$ and important phase information
that allows the translation of the $x', y'$ measurements in \prt{D\to
  K\pi} into the unprimed parameters. And finally, in the decay chain
\prt{B^\pm \to D^0 K^{\pm}}, and equivalently \prt{B^0 \to D^0 K^*},
phase information from the B decay is encoded in the \prt{D} Dalitz
plot, allowing a theoretically clean measurement of $\gamma$. This
method provides currently the best constraint on $\gamma$. Input from
the analysis of quantum-correlated \prt{D\bar{D}} pairs at CLEO-c will
significantly reduce the model uncertainty in this group of
measurements, in the \prt{D^0 \to K_S \pi \pi} mode as well as other
$2, 3$ and 4-body decay modes of the \prt{D^0}. This is of particular
importance for future facilities such as the proposed Super Flavour
Factory, and LHCb, which is due to start data taking in 2009, and
where these uncertainties would soon be the limiting factor in the
precision on $\gamma$.

\end{document}